\documentclass[aps,prl,twocolumn,notitlepage,superscriptaddress]{revtex4-1}
\usepackage[colorlinks=true, citecolor=magenta, linkcolor=blue, urlcolor=blue]{hyperref}
\usepackage{graphicx}
\usepackage{amsmath}
\usepackage{amssymb}
\usepackage{amsfonts}
\usepackage{xcolor}
\usepackage{tabularx,booktabs}
\usepackage{multirow}
\usepackage{braket}
\usepackage{bm}      
\usepackage[symbol]{footmisc}
\usepackage{tikz}
\usetikzlibrary{shapes.geometric, arrows}
\usetikzlibrary{decorations.markings}

\allowdisplaybreaks

\begin{document}
\title{Anyonic Chern insulator in graphene induced by surface electromagnon vacuum fluctuations}
\author{Xinle Cheng}
\affiliation{Max Planck Institute for the Structure and Dynamics of Matter, Luruper Chaussee 149, 22761 Hamburg, Germany}

\author{Emil {Vi\~nas Bostr\"om}}
\affiliation{Max Planck Institute for the Structure and Dynamics of Matter, Luruper Chaussee 149, 22761 Hamburg, Germany}

\author{Frank Y. Gao}
\affiliation{Department of Physics, The University of Texas at Austin, Austin, Texas, 78712}

\author{Edoardo Baldini}
\affiliation{Department of Physics, The University of Texas at Austin, Austin, Texas, 78712}

\author{Dante M. Kennes}
\affiliation{Institute for Theory of Statistical Physics, RWTH Aachen University, 52056 Aachen, Germany}
\affiliation{Max Planck Institute for the Structure and Dynamics of Matter, Luruper Chaussee 149, 22761 Hamburg, Germany}

\author{Angel Rubio}
\affiliation{Max Planck Institute for the Structure and Dynamics of Matter, Luruper Chaussee 149, 22761 Hamburg, Germany}
\affiliation{Initiative for Computational Catalysis, The Flatiron Institute, Simons Foundation, New York City, NY 10010, USA}

\date{\today}

\begin{abstract}
Sub-wavelength cavities have emerged as a promising platform to realize strong light-matter coupling in condensed matter systems. Previous studies are limited to dielectric sub-wavelength cavities, which preserve time-reversal symmetry. Here, we lift this constraint by proposing a cavity system based on magneto-electric materials, which host surface electromagnons with non-orthogonal electric field and magnetic field components. The quantum fluctuations of the surface electromagnons drive a nearby graphene monolayer into an anyonic Chern insulator, characterized by anyonic quasi-particles and a topological gap that decays polynomially with the graphene-substrate distance. Our work opens a path to controllably break time-reversal symmetry and induce exotic quantum states through cavity vacuum fluctuations.

\end{abstract}

\maketitle

\textit{Introduction}.-- Recently, the regime of strong light-matter coupling has been realized in condensed matter systems~\cite{li2018vacuum,bloch2022strongly,frisk2019ultrastrong,wang2013observation,de2021colloquium,zhou2023pseudospin,10.1063/5.0083825,bloch2022strongly}. In such systems, light serves not only as a probe of near-equilibrium responses, but also as an active means to alter material properties and induce new phases of matter~\cite{bao2022light,chattopadhyay2025metastable, fan2024chiral,truc2023light,vinas2023controlling}. An important milestone in this direction was the observation of quantized transport in monolayer graphene, driven by strong circularly polarized MIR light~\cite{McIver2019}, signaling a light-induced topological phase transition. Strong driving, however, produces significant heating~\cite{mallayya2019heating, de2021colloquium}, and complicates both experiments and the theoretical understanding. In addition, field-induced effects are intrinsically transient~\cite{ito2023build} and decay quickly after the pulse, whereas many applications require long-lived or stable modifications.

\begin{figure}[b]
    \centering
    \includegraphics[width=1.0\columnwidth]{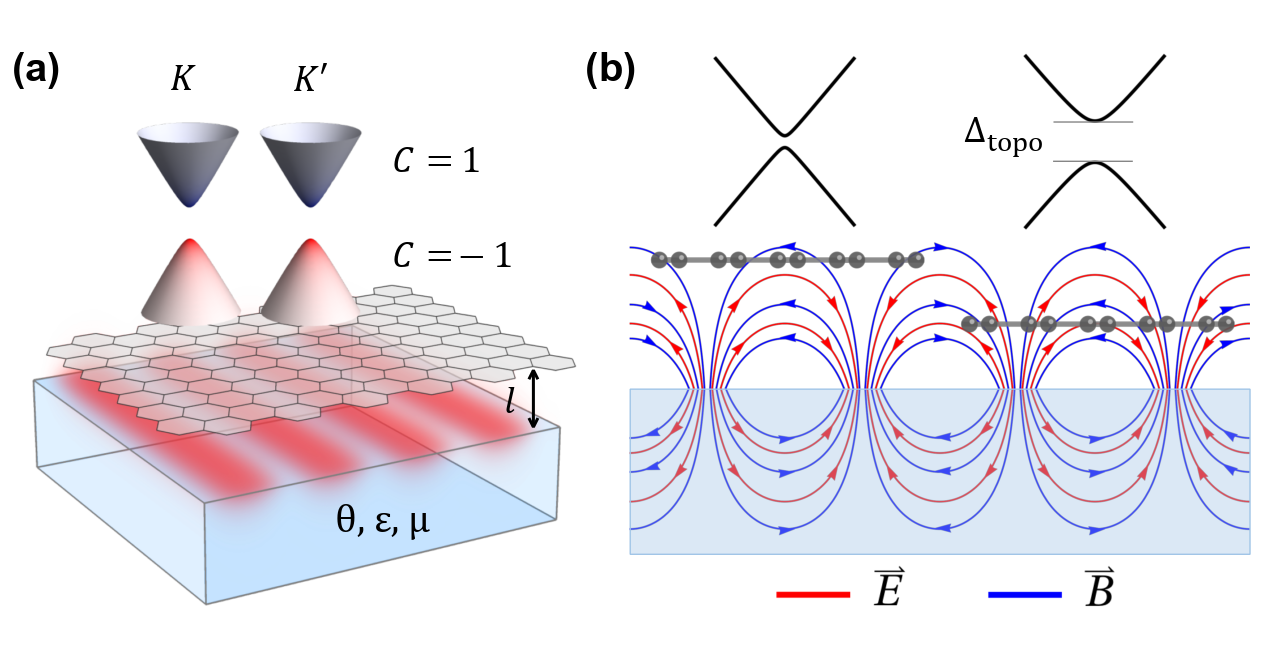}
    \caption{(a) Schematic of a magneto-electric sub-wavelength cavity: monolayer graphene sits a distance $l$ above a magneto-electric substrate. Coupling to surface electromagnons opens a topological gap $\Delta_\mathrm{topo}$ at the Dirac points, driving the graphene layer into a Chern insulating state. (b) Electric (red) and magnetic (blue) field profiles for a surface electromagnon mode. The fields are parallel inside the substrate and anti-parallel in vacuum. Furthermore, the fields are confined near the surface, consequently $\Delta_\mathrm{topo}$ grows as $l$ decreases.}
    \label{fig:setup}
\end{figure}

Cavity material engineering addresses these issues by utilizing tailored electromagnetic vacuum fluctuations to modify materials in equilibrium~\cite{enkner2025tunable,appugliese2022breakdown,lu2025cavity,ashida2023cavity,hubener2024quantum,ruggenthaler2014quantum,10.1063/5.0083825, garcia2021manipulating}. Recently, sub-wavelength cavities has emerged as a promising architecture for cavity engineering~\cite{strupiechonski2013hybrid, herzig2024high}, where a target system is placed near the surface of a dielectric substrate ( Fig.~\ref{fig:setup}). The substrate hosts confined electromagnetic modes, such as plasmons or phonon-polaritons, localized at the interface but with electric fields extending some distance into the vacuum. These modes overcome the diffraction limit, and enable strong light-matter interaction with proximate materials~\cite{todorov2009strong, feres2021sub,kipp2024cavity}. 

To date, sub-wavelength cavities have been proposed and realized as a means to modify structural transitions, magnetism and superconductivity~\cite{keren2025cavity,bostrom2024equilibrium,thomas2021large,sentef2018cavity,weber2023cavity}. These proposals exploit the large longitudinal electric field component of the surface modes, while assuming that the magnetic field component is negligible. This hinders their application to phenomena where time-reversal symmetry breaking plays a prominent role, such as many topological and magnetic effects. It is however possible, as we show below, to realize surface modes with a significant magnetic field component. Because their field distribution breaks time-reversal symmetry, these modes yield qualitatively new effects compared to conventional dielectric polaritons.

Here, we extend the functionality of sub-wavelength cavities to time-reversal symmetry breaking phenomena, by considering a magneto-electric substrate that hosts surface modes with both electric and magnetic characters. These modes, often denoted electromagnons, are ubiquitous in materials with significant magneto-electric coupling, such as multiferroics~\cite{pimenov2006possible, gao2024giant, kurumaji2017electromagnon}. We analytically quantize the electromagnon surface modes in the deep sub-wavelength regime, and study their interaction with a graphene monolayer positioned a distance $l$ above the substrate (Fig.~\ref{fig:setup}).  We show that this interaction leads to an effective magnetic flux attachment for external charges, such that the graphene quasi-particles acquire anyon exchange statistics. Finally, we demonstrate that the quantum fluctuations of the surface modes transform graphene into a Chern insulator, with a topological gap up to $\sim1$ K.


\textit{Surface electromagnon modes}.-- We first consider an interface between vacuum ($z>0$) and a homogeneous, isotropic magneto-electric material ($z<0$). The material is characterized by the constitutive relations
\begin{subequations}\label{eq2}
    \begin{align}
    &\bm{D} = \varepsilon\bm{E} +\theta \bm{B} \label{eq2_1},\\
    &\bm{H} = -\theta\bm{E} + \frac{1}{\mu} \bm{B} \label{eq2_2},
    \end{align}
\end{subequations}
where $\bm{D}$ is the displacement field, $\bm{E}$ the electric field, $\bm{B}$ the magnetic induction, and $\bm{H}$ the magnetizing field. The functions $\varepsilon$, $\mu$, and $\theta$ denote the frequency dependent electric permittivity, magnetic permeability and magneto-electric coupling, respectively. Throughout, we set the relativistic units $\varepsilon_0 = \mu_0 = c = \hbar = 1$, and use $\alpha \approx 1/137$ to denote the fine structure constant. 

We seek solutions of Maxwell's equation that are localized to the interface, which can be labeled by an in-plane momentum $\bm{q}$ due to translational symmetry. The out-of-plane component of the reciprocal vector is imaginary, and will be denoted $\kappa$ and $\kappa'$ on the material and vacuum sides. Boundary conditions together with bulk solutions yield the conditions $q^2 + \kappa^2 = \varepsilon\mu\omega^2$, $q^2 + \kappa'^2 = \omega^2$, and 
\begin{equation}
    (\varepsilon\kappa'-\kappa)(\mu\kappa'-\kappa) - \theta^2\mu\kappa\kappa' = 0,
    \label{master_eq}
\end{equation}
which determine the surface mode dispersion $\omega_{s,\bm{q}}$~\cite{Supplemental}.

In realistic systems, the deep sub-wavelength limit $\nu = \omega_{s,\bm{q}}/q \ll 1$ holds over almost the entire the Brillouin zone, except near $q = 0$~\cite{Supplemental}. We therefore adopt this limit henceforth. Within the deep sub-wavelength limit, $\kappa = -iq + O(\nu^2)$ and $\kappa' = iq + O(\nu^2)$, where the opposite signs ensure that the surface modes decay away from the interface. Combining this result with Eq.~(\ref{master_eq}) shows that for large $q$ the surface mode dispersion approaches a constant, $\omega_s$, fixed by
\begin{equation}
 (\varepsilon_s+1)(\mu_s+1) + \theta_s^2\mu_s = 0.
\label{eq_surface_mode_frequency}
\end{equation}
Here $\varepsilon_s$, $\mu_s$ and $\theta_s$ are the material properties evaluated at $\omega_s$. The corresponding eigenmodes of scalar potential $\phi_{\bm{q}}$ and vector potentials $\bm{A}_{\bm{q}}$, in the Coulomb gauge, read
\begin{equation}
    \begin{aligned}
        &\phi_{\bm{q}} = \mathcal{N}_q\, e^{i\bm{q}\cdot\bm{r}_{\parallel} \pm qz},\\
        &\bm{A}_{\bm{q}} = i\mathcal{N}_q\tilde{\theta}_s(\hat{\bm{z}}\times\hat{\bm{q}})e^{i\bm{q}\cdot\bm{r}_{\parallel} \pm qz}.
        \label{eq_eigenmode_A_phi}
    \end{aligned}
\end{equation}
Here $\bm{r}_{\parallel}$ is the in-plane coordinate, $\tilde{\theta}_s = \mu_s\theta_s/(\mu_s+1)$, and $\pm$ correspond to $z<0$ and $z>0$. $\mathcal{N}_q$ is a normalization constant to be fixed during the quantization procedure. Figure~\ref{fig:setup}(b) shows the resulting field profiles, and that $\bm{E}$ and $\bm{B}$ are parallel inside the substrate but anti-parallel in the vacuum outside. The electromagnon surface modes thereby break time-reversal symmetry and mirror symmetry perpendicular to the interface.


\textit{Quantizing the surface modes}.--
Having derived the classical solutions of Maxwell's equations, we now proceed to quantize the surface modes. For dispersive dielectric medium, it is known that a self-consistent quantization must include the underlying matter excitations~\cite{gubbin2016real}. However, a corresponding scheme for magneto-electric medium is still missing. Here we provide such a scheme under the deep sub-wavelength limit. We denote the canonical coordinate of the electromagnon by a vector field $\bm{X}(\bm{r})$, which corresponds to the underlying lattice distortion or spin fluctuation that drives the multiferroic response. Utilizing the isotropic nature of the medium, the contribution of the electromagnon to the polarization $\bm{P}$ and the magnetization $\bm{M}$ can be written as $\bm{P} = \omega_p\bm{X}$ and $\bm{M} = \omega_m\bm{X}$, respectively. Here $\omega_p$ and $\omega_m$ have units of energy, and are proportional to the electric and magnetic dipoles of the electromagnon excitation. In the deep sub-wavelength limit $\omega_s/q \ll 1$, the photon spectral weight vanishes from the surface modes, and the quantization of $\bm{X}$ is 
\begin{equation}
\begin{aligned}
    &\bm{X}(\bm{r}) = \sum_{\bm{q}} \frac{1}{\sqrt{2\omega_s}}\left[\bm{X}_{\bm{q}}(\bm{r}) a_{\bm{q}} + \bm{X}^*_{\bm{q}}(\bm{r})a^\dagger_{\bm{q}}\right],
\end{aligned}
\end{equation}
where the normal modes $\bm{X}_{\bm{q}}$ satisfy the normalization condition $\int\mathrm{d}\bm{r}|\bm{X}_{\bm{q}}(\bm{r})|^2 = 1$. From Eq.~(\ref{eq2_1}), the polarization $\bm{P}$ can be expressed as $\bm{P} = (\varepsilon-1)\bm{E} + \theta\bm{B}$, which allows us to rewrite the normalization of surface modes in terms of $\bm{E}$ and $\bm{B}$ as
\begin{equation}
   \frac{1}{\omega_p^2}\int_{z<0}\mathrm{d}\bm{r}\left|(\varepsilon_s-1)\bm{E}_\lambda(\bm{r})+\theta_s\bm{B}_\lambda(\bm{r})\right|^2 = 1.
\end{equation}
This relation, together with Eqs.~(\ref{eq_surface_mode_frequency}), determines the normalization constant $\mathcal{N}_q = \omega_p/2\sqrt{q S}$,
where $S$ is the interface surface area. The quantized scalar and vector potentials then read, in the Coulomb gauge:
\begin{equation}
\begin{aligned}
    &\phi_{\bm{r}} = \sum_{\bm{q}}\frac{1}{\sqrt{2\omega_s}}\left(\phi_{\bm{q}}a_{\bm{q}} + \phi^*_{\bm{q}}a^\dagger_{\bm{q}}\right), \\
    &\bm{A}_{\bm{r}} = \sum_{\bm{q}}\frac{1}{\sqrt{2\omega_s}}\left(\bm{A}_{\bm{q}}a_{\bm{q}} + \bm{A}^*_{\bm{q}}a^\dagger_{\bm{q}}\right).
\end{aligned}
\end{equation}
In the above derivation, we retain an arbitrary frequency dependence of $\varepsilon$, $\mu$ and $\theta$, and exploit the deep sub-wavelength limit. In the Supplementary Material (SM) we present an alternative approach, where a Lorentzian and predominantly dielectric excitation is assumed. This approach shows that the photon spectrum weight is suppressed by a factor of $\omega_p^2/q^2$ and vanishes in the deep sub-wavelength limit, reducing to the above quantization scheme.  

\textit{Interaction with graphene}.-- Having quantized the surface mode, we now examine its impact on a nearby graphene sheet (Fig.~\ref{fig:setup}). In graphene spin-orbit coupling is negligible, so spin and charge decouple. For a distance $l$  between graphene and the magneto-electric substrate much larger than the lattice spacing, each valley can be treated independently as a two-dimensional Weyl fermion, coupled to the surface mode via minimal coupling~\cite{Supplemental}. The resulting Hamiltonian reads
\begin{equation}
\begin{aligned}
    H &= \int\mathrm{d}\bm{r}\,\bar{\psi}_{\bm{r}}\left[v_F \bm{\sigma}\cdot(-i\nabla + e\bm{A}_{\bm{r}}) -e\phi_{\bm{r}}\right]\psi_{\bm{r}}\\
    &\quad+\sum_{\bm{q}}\omega_s \Big(b_{\bm{q}}^\dagger b_{\bm{q}} + \frac{1}{2} \Big)\\
    &= \sum_{\bm{q}}\omega_s \Big(b_{\bm{q}}^\dagger b_{\bm{q}} + \frac{1}{2} \Big) + \sum_{\bm{k}}v_F\bar{\psi}_{\bm{k}}(\bm{\sigma}\cdot\bm{k})\psi_{\bm{k}}\\
    &+\frac{1}{\sqrt{S}}\sum_{\bm{k},\bm{q}}\bar{\psi}_{\bm{k+q}}(i\mathcal{A}_{\bm{q}}\cdot\bm{\sigma} - \Phi_{\bm{q}})\psi_{\bm{k}}b_{\bm{q}} + H.c.
\end{aligned}
\label{eq_int_ham}
\end{equation}
where $\psi_{\bm{r}}$ is a two-component spinor in the sub-lattice basis, $v_F$ is the graphene Fermi velocity, and $\bm{\sigma}$ is the vector of Pauli matrices. Here, for notational simplicity, we define
\begin{equation}
\begin{aligned}
    &\mathcal{A}_{\bm{q}}=\frac{ev_F\tilde{\theta}_s\omega_p}{2} \sqrt{\frac{q}{2\omega_s}}\frac{e^{-ql}}{q}(\hat{\bm{z}}\times\hat{\bm{q}}),\\
    &\Phi_{\bm{q}}=\frac{e\omega_p}{2} \sqrt{\frac{q}{2\omega_s}}\frac{e^{-ql}}{q}.
\end{aligned}
\end{equation}


\textit{Flux attachment}.-- We first analyze $H$ via a Schrieffer-Wolff transformation~\cite{bravyi2011schrieffer,Supplemental}, which when the electronic kinetic energy is negligible eliminates the coupling between the electrons and the surface mode. The resulting Hamiltonian contains induced density-density, current-current, and density-current interactions between electrons at $\bm{r}$ and $\bm{r}'$. Their long-distance forms are
\begin{equation}
\begin{aligned}
    &V_{dd} = -\frac{1}{16\pi}\frac{\omega_p^2}{\omega_s^2}\int\mathrm{d}\bm{r}\int\mathrm{d}\bm{r}'\frac{\rho_{\bm{r}}\rho_{\bm{r}'}}{|\bm{r}-\bm{r}'|}, \\
    &V_{cc} = -\frac{\tilde{\theta}_s^2}{16\pi}\frac{\omega_p^2}{\omega_s^2}\int\mathrm{d}\bm{r}\int\mathrm{d}\bm{r}'\frac{1}{|\bm{r}-\bm{r}'|}(\bm{J}_{\bm{r}}\cdot\hat{\bm{\eta}})(\bm{J}_{\bm{r}'}\cdot\hat{\bm{\eta}}), \\
    &V_{dc} = -\frac{\tilde{\theta}_s}{8\pi}\frac{\omega_p^2}{\omega_s^2}\int\mathrm{d}\bm{r}\int\mathrm{d}\bm{r}' \frac{\rho_{\bm{r}'}}{|\bm{r}-\bm{r}'|} (\hat{\bm{z}}\times\hat{\bm{\eta}}) \cdot \bm{J}_{\bm{r}},
\end{aligned}
\end{equation}
where $\rho_{\bm{r}} = -e\bar{\psi}_{\bm{r}}\psi_{\bm{r}}$ and $\bm{J}_{\bm{r}} = -e v_F\bar{\psi}_{\bm{r}}\bm{\sigma}\psi_{\bm{r}}$ are density and current operators, respectively, and $\hat{\bm{\eta}} = (\bm{r}-\bm{r}')/|\bm{r}-\bm{r}'|$ is a unit vector pointing from $\bm{r}$ to $\bm{r}'$.

These interactions can be interpreted as follows: The density-density term corresponds to a dielectric screening from the substrate, whereas the Amperean current-current term gives an attractive (repulsive) force along the line connecting two co-propagating (contra-propagating) electrons. More intriguing is the density-current interaction, which can be interpreted as a magnetic flux of magnitude $(\alpha\tilde{\theta}_s/2) (\omega_p^2/\omega_s^2)\Phi_0$ being attached to each electron, with $\Phi_0$ the flux quantum. This mechanism bears striking similarities to the composite-fermion physics of the fractional quantum Hall liquid~\cite{tong2016lectures}, with the difference that here the attached flux is only a small fraction of $\Phi_0$. 

An important implication of the flux attachment is its influence on the exchange statistics of graphene quasi-particles. To calculate its effect, we eliminate $V_{dc}$ by the unitary transformation 

\begin{equation}
    \psi_{\bm{r}}\rightarrow \psi'_{\bm{r}} = \exp\left[{2 i\pi f \int \mathrm{d}\bm{r}'\Theta(\bm{r},\bm{r}')\rho_{\bm{r}'}}\right]\psi_{\bm{r}},
    \label{eq_comp_fermion}
\end{equation}
where $f = (\alpha\tilde{\theta}_s/2) (\omega_p^2/\omega_s^2)$ and $\Theta(\bm{r},\bm{r}')$ is a multi-valued function satisfying $\nabla_{\bm{r}}\Theta(\bm{r},\bm{r}') = \hat{\bm{z}}\times (\bm{r}-\bm{r}')/|\bm{r}-\bm{r}'|^2$. The resulting commutation relation is
\begin{equation}
    \psi'_{\bm{r}}\psi'_{\bm{r}'} + e^{i\pi f}\psi'_{\bm{r}'}\psi'_{\bm{r}} = 0,
\end{equation}
and shows that the quasi-electrons obey Abelian anyon statistics with an anyon angle of $ (1+f)\pi$~\cite{wilczek1982quantum}. This cavity-induced anyon statistics can in principle be detected by quasi-particle interferometry~\cite{carrega2021anyons, samuelson2024anyonic}.


\begin{figure*}[t]
    \centering
    \includegraphics[width=1.8\columnwidth]{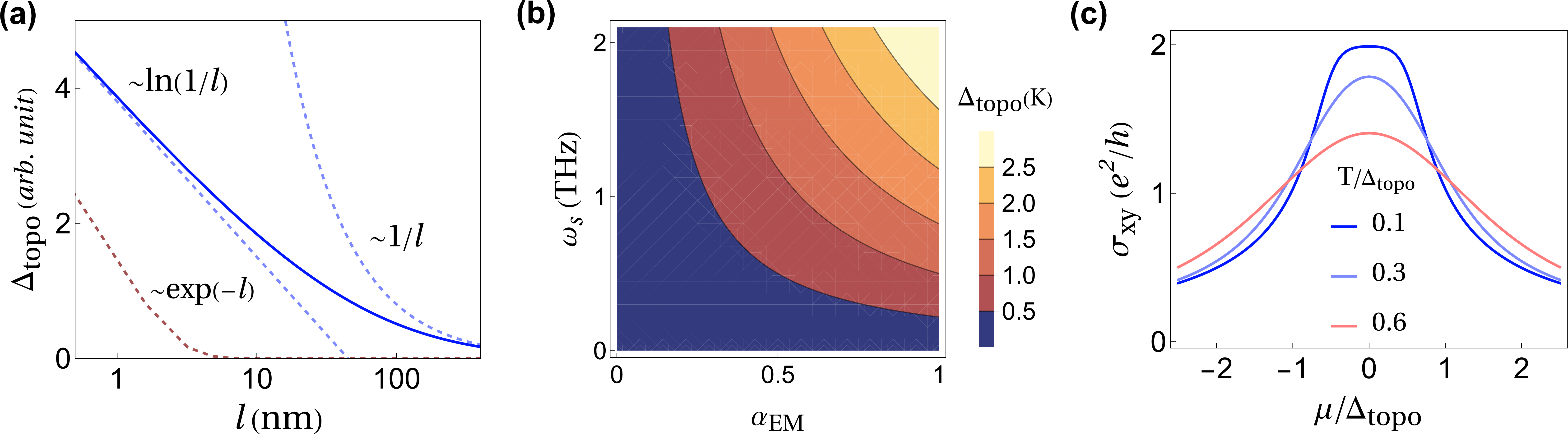}
    \caption{(a) Scaling of the topological gap $\Delta_\mathrm{topo}$ with the graphene-substrate distance $l$, for $v_F = 10^6$ m/s and $\omega_s = 1$ THz. The solid blue line shows the exact result of Eq.~(\ref{eq_mz}), the dotted blue lines show the $l\rightarrow 0$ and $l\rightarrow \infty$ asymptotic forms discussed in the main text, and the dotted red line illustrates the expected behavior of a proximity-induced effect with exponential decay. (b) Magnitude of the topological gap $\Delta_\mathrm{topo}$ as a function of $\alpha_\mathrm{EM}$ and the surface electromagnon frequency $\omega_s$, with $v_F=10^6$ m/s and $l=3\mathrm{\AA}$. 
    (c) Single-particle Hall conductance $\sigma_{xy}$ as a function of chemical potential $\mu$ at various temperatures.}
    \label{fig:delta_topo}
\end{figure*}

\textit{Topological phase transition}.-- To analyze how the surface modes affect the graphene band topology, we calculate the electronic self-energy. To second order in the electron-electromagnon coupling, the self-energy at frequency $\omega$ and momentum $\bm{k}$ reads
\begin{equation}
\begin{aligned}
 \Sigma(\omega,\bm{k}) &= \int \frac{\mathrm{d}\bm{q}\mathrm{d}\omega_b}{(2\pi)^3} \frac{1}{i\omega_b - \omega_s} (i\mathcal{A}_{\bm{q}}\cdot\bm{\sigma} + \Phi_{\bm{q}}) \\
 &\hspace*{0.8cm} \times G_f(\omega_b+\omega,\bm{k}+\bm{q}) (i\mathcal{A}_{\bm{q}}\cdot\bm{\sigma} - \Phi_{\bm{q}}) \\
 &- \int \frac{\mathrm{d}\bm{q}\mathrm{d}\omega_b}{(2\pi)^3} \frac{1}{i\omega_b + \omega_s} (i\mathcal{A}_{\bm{q}}\cdot\bm{\sigma} - \Phi_{\bm{q}}) \\
 &\hspace*{0.8cm}\times G_f(\omega_b+\omega,\bm{k}-\bm{q}) (i\mathcal{A}_{\bm{q}}\cdot\bm{\sigma} + \Phi_{\bm{q}}).
\end{aligned}
\end{equation}
Here $G_f(\omega,\bm{k})$ is the imaginary time non-interacting fermion Green's function, which is defined as
\begin{equation}
    G_f(\omega,\bm{k}) = -\frac{i\omega + v_F\bm{k}\cdot\bm{\sigma}}{\omega^2 + v_F^2 k^2}.
\end{equation}
Focusing on the possible opening of a gap at the Dirac point, we take $\omega$ and $\bm{k}$ to be zero, and find that the $\sigma_z$ component of the self-energy is
\begin{equation}
\begin{aligned}
\Sigma_z(0,\bm{0})  = i[\sigma_x, \sigma_y]\int\frac{\mathrm{d}\bm{q}}{(2\pi)^2}\frac{(\hat{\bm{q}}\times\mathcal{A}_{\bm{q}})\Phi_{\bm{q}}}{\omega_s+v_F q}.
\end{aligned}
\label{eq_Sigma_z}
\end{equation}
Since $G_f$ is diagonal in spin and valley indexes, each sector can be treated independently. For the valley of opposite chirality, the corresponding result is obtained by replacing $\bm{\sigma}$ by $\tilde{\bm{\sigma}} = (\sigma_x, -\sigma_y)$~\cite{Supplemental}, which flips the sign of the commutator. We therefore obtain the self-energy $\Sigma_z = \pm m\sigma_z$ for the two valleys, with
\begin{equation}
    m = 2\int\frac{\mathrm{d}\bm{q}}{(2\pi)^2}\frac{(\hat{\bm{q}}\times\mathcal{A}_{\bm{q}})\Phi_{\bm{q}}}{\omega_s+v_F q}.
    \label{eq_mz}
\end{equation}
Each valley is found to carry a $C = -1/2$ Chern number ~\cite{Supplemental}, such that combing the two valleys gives a $C = -1$ Chern insulator (for each spin), with a topological gap $\Delta_\mathrm{topo} = 2|m|$.  

Figure~\ref{fig:delta_topo} shows the evolution of the topological gap with the graphene-substrate distance $l$. For small separations $l \ll v_F/\omega_s$, the asymptotic form is $\Delta_\mathrm{topo} \sim \alpha\tilde{\theta}_s (\omega_p^2/\omega_s) [ \ln(v_F/2\omega_s l) - \gamma_E ]$ with $\gamma_E$ Euler's constant. This expression is dominated by the logarithmic term, leading to a logarithmic increase of the gap at small distances. In contrast, for large separations $l \gg v_F/\omega_s$, the asymptotic form is $\Delta_\mathrm{topo} \sim (\alpha\tilde{\theta}_s/2) (\omega_p^2/\omega_s^2) (v_F/l)$. 
The polynomial decay with $l$ is a hallmark of long-range interactions, and distinguishes the surface mode contribution from possible proximity-induced effects. Fig.~\ref{fig:delta_topo} also shows the single-particle Hall conductance as a function of chemical potential $\mu$~\cite{Supplemental}. For low temperatures $T \ll \Delta_\mathrm{topo}$, $\sigma_{xy} = 2e^2/h$ when $\mu$ lies inside the gap and the system exhibits an anomalous quantum Hall effect. For temperatures comparable to the gap, the quantized conductance is washed out.

Finally, we note that while the self-energy is generally gauge dependent, the $\sigma_z$ component induced by the surface mode is gauge invariant~\cite{Supplemental}. This is consistent with the topological gap being a physical observable.


\textit{Implications for real materials}.-- To connect the parameters of our model to properties of real materials, we introduce a phenomenological model for magneto-electric media. We treat $\bm{X}$ as a local harmonic oscillator coupled to external electric and magnetic fields through the equation of motion
$\ddot{\bm{X}} + \omega^2_o \bm{X} - \omega_p \bm{E} - \omega_m \bm{H}  =0$.
The electric polarization and magnetization are determined through $\bm{P} = \omega_p \bm{X}$ and $\bm{M} = \omega_m \bm{X}$, showing that the model is well suited for electromagnons carrying both electric and magnetic dipole moments~\cite{gao2024giant}. From this parameterization the response functions are computed to be $\varepsilon = 1 - \omega_p^2/(\omega^2 - \omega_o^2 - \omega_m^2)$, $\mu^{-1} = 1 + \omega_m^2/(\omega^2 - \omega_o^2 - \omega_m^2)$ and $\theta = -\omega_p\omega_m/(\omega^2 - \omega_o^2 - \omega_m^2)$.
Using these expressions in Eq.~(\ref{eq_surface_mode_frequency}) determines the frequency $\omega_s$ for large momentum surface modes to be
\begin{equation}
    \omega_s = \sqrt{\omega_o^2+\frac{1}{2}(\omega_m^2+\omega_p^2)}.
\end{equation}
The frequency satisfies $\tilde{\theta}_s \omega_p^2/\omega_s^2 = -\omega_p\omega_m/\omega_s^2$, and we define $\alpha_\mathrm{EM} = \omega_p\omega_m/\omega_s^2$, which is the contribution to the zero frequency magneto-electric coupling from the electromagnons. Note that $\alpha_\mathrm{EM}$ is in general different from the static magneto-electric coupling of a material, which also has contributions from mechanisms such as domain alignment. The topological gap increases monotonically with $\alpha_\mathrm{EM}$ and $\omega_s$, as illustrated in Fig.~\ref{fig:delta_topo}(b). 

As a candidate substrate material, we consider the hexaferrite $\mathrm{Ba_2 Mg_2 Fe_{12} O_{22}}$, which is known to host a bulk electromagnon with frequency $\omega_o\approx 0.7$ THz~\cite{kida2011gigantic}. By adding a damping to $\varepsilon(\omega)$ and fitting the resulting Lorentzian to THz absorption measurements~\cite{kida2009electric}, we find $\omega_p = 0.96$ THz. Due to the lack of measured dynamical magneto-electric couplings, we resort to the order-of-magnitude estimate $\omega_m \sim \mu_B \sqrt{2S\mu_0\hbar\omega_o/V_c}$~\cite{Supplemental}. Using $S = 5/2$ for the spin of the magnetic $\mathrm{Fe}^{3+}$ ion, and $V_c \approx 108$~{\AA}$^3$ for the average volume per spin, the resulting coupling is $\omega_m \sim 0.1$ THz.

For these values of $\omega_o$, $\omega_p$ and $\omega_m$, $\alpha_\mathrm{EM} \approx 0.1$, and taking $v_F = 10^6$ m/s and $l = 3$~{\AA} gives a gap of $\Delta_\mathrm{topo} \approx 0.2$~K. However, the value of topological gap can be further increased by material design. Specifically, for a material near a multiferroic phase transition, the electromagnon mode will soften so that $\omega_o\ll \omega_p,\omega_m$. The coupling $\alpha_{\text{EM}}$ then takes a maximal value of 1 when $\omega_m = \omega_p$, and the optimal topological gap is $\Delta_\mathrm{opt} = \alpha\omega_s\left[\ln(v_F/2\omega_s l)-\gamma_E\right]$. Taking $\omega_m = \omega_p = 1~\text{THz}$, and the same values for $v_F$ and $l$ as above, we find $\Delta_\text{opt} \approx 1.8~\text{K}$. 

\textit{Discussions}.-- To highlight the novelty of our study, we compare our proposed magneto-electric surface cavity to the previously studied cavities that also induce symmetry breaking, such as chiral cavities~\cite{hubener2021engineering, tay2025terahertz, wang2019cavity,voronin2022single}. In our setup, the finite in-plane momentum of the surface modes induces a non-trivial spatial correlation between the electrons [see Eq.~\ref{eq_comp_fermion}], and is the key behind the anyonic statistics for the quasi-particles. Contrarily, in the case of a chiral cavity, because of the vanishing momentum of the photon modes, there is no flux attachment and the quasi-particles remain fermionic~\cite{liu2025cavity}. We also alter our setup slightly, considering a geometry with two magneto-electric interfaces~\cite{Supplemental}, and we find that the modes with zero in-plane momentum realize a chiral cavity. Our proposal therefore defines a concrete yet more versatile framework for symmetry-breaking cavities, making full use of multiple modes and finite momenta. To distinguish the cavity system in this study, and as the electric and magnetic components of the cavity modes are non-orthogonal, we name it an "axionic cavity".

\textit{Conclusions}.-- We have demonstrated that magneto-electric materials can host sub-wavelength confined surface modes which break time-reversal symmetry. Within a general framework based on the deep sub-wavelength approximation, we derived and quantized the surface electromagnon modes. By coupling them to a monolayer graphene, we further developed an effective flux attachment picture for the anyonic quasi-electrons, and demonstrated that the quantum fluctuations of the surface electromagnons can drive graphene into a Chern insulating state. The resulting topological gap exhibits a polynomial scaling with the graphene-substrate distance, in contrast to conventional exponential decay in interfacial proximity effects. A gap size in the range of $0.1 - 1$ K is estimated based on realistic material parameters. Importantly, our framework is largely insensitive to microscopic details, making it applicable to a broad class of magneto-electric materials. 

Our study lays the groundwork for a new class of symmetry-breaking cavity systems that we coin axionic cavities. Future work should investigate the interplay between the surface electromagnon mode and the chiral edge states of the induced Chern insulator~\cite{appugliese2022breakdown}. Another intriguing avenue is to explore the regime where the chemical potential lies outside the topological gap. In this case the effective flux attachment is reminiscent of the fractional quantum Hall effect, but non-integer flux values can be realized. 
More broadly, our work presents a realistic route to break time-reversal symmetry and induce exotic phases of matter with quantum electromagnetic environment, for systems in thermal equilibrium.

\begin{acknowledgments}
 \textit{Acknowledgments}.-- We acknowledge stimulating discussions with Marios H. Michael. We thank Zhiyang Zheng for helpful feedback on the manuscirpt. X.C., E.V.B. and A.R. acknowledge support from the Cluster of Excellence “CUI: Advanced Imaging of Matter”–EXC 2056–project ID 390715994, SFB-925 “Light induced dynamics and control of correlated quantum systems”–project ID 170620586 of the Deutsche Forschungsgemeinschaft (DFG), the European Research Council (ERC-2024-SyG-UnMySt–101167294), and the Max Planck-New York City Center for Non-Equilibrium Quantum Phenomena. D.M.K. acknowledges support by the DFG via Germany's Excellence Strategy-Cluster of Excellence Matter and Light for Quantum Computing (ML4Q, Project No. EXC 2004/1, Grant No.390534769) and individual Grant No. 508440990. Work at UT Austin was primarily supported by the W. M. Keck Foundation under grant 996588 (F.Y.G.) and the Army Research Office under grant W911NF-23-1-0394 (E.B.). F.Y.G. acknowledges additional support from the Texas Quantum Institute. The Flatiron Institute is a division of the Simons Foundation.
\end{acknowledgments}

\bibliography{references}

\end{document}